\def\apj{{ApJ}}
\def\bd{BD+17\deg 3248}
\def\cs{\mbox{CS~22892-052}}
\def\deg{{$^{\circ}$}}

\def\cm2{cm$^{-2}$}

\def\c2{C~{\sc ii}}
\def\c4{C~{\sc iv}}
\def\fe2{Fe~{\sc ii}}
\def\fe3{Fe~{\sc iii}}
\def\mg1{Mg~{\sc i}}
\def\mg2{Mg~{\sc ii}}
\def\si2{Si~{\sc ii}}
\def\si4{Si~{\sc iv}}
\def\al2{Al~{\sc ii}}
\def\al3{Al~{\sc iii}}
\def\o1{O~{\sc i}}
\def\n1{N~{\sc i}}
\def\h1{H~{\sc i}}

\def\approxlt{\mathrel{\spose{\lower 3pt\hbox{$\sim$}}
        \raise 2.0pt\hbox{$<$}}}
\def\approxgt{\mathrel{\spose{\lower 3pt\hbox{$\sim$}}
        \raise 2.0pt\hbox{$>$}}}

\documentclass[article]{gwp80}
\usepackage{graphicx}
\usepackage{amssymb}
\tabletypesize{\normalsize}  

\def\plotone#1{\centering \leavevmode
\includegraphics[width=.95\columnwidth]{#1}}

\def\plotone#1{\centering \leavevmode
\includegraphics[width=.95\columnwidth]{#1}}

\shortauthors{Cowan et al.}
\shorttitle{$r$-Process Abundance Signatures}
\begin{document}
\large
\pagenumbering{arabic}
\setcounter{page}{1}

\title{$r$-Process Abundance Signatures in Metal-Poor Halo Stars}

%
%
\author{{\noindent John J. Cowan{$^{\rm 1}$}, Ian U. Roederer{$^{\rm 2}$}, 
Christopher Sneden{$^{\rm 3}$}
and James E. Lawler{$^{\rm 4}$}\\
\\
{\it (1) HLD Department of Physics \& Astronomy, University of Oklahoma, 
Norman, OK, USA\\
(2) Carnegie Observatories, Pasadena, CA, USA\\ 
(3) Department of Astronomy and McDonald Observatory,
University of Texas, Austin, TX, USA\\ 
(4) Department of Physics, University of Wisconsin,
Madison, WI, USA} 
}
}
%
\email{(1) cowan@nhn.ou.edu (2) iur@obs.carnegiescience.edu
(3) chris@verdi.as.utexas.edu\\ (4) jelawler@wisc.edu}

\begin{abstract}
Abundance observations indicate the presence of
rapid-neutron capture ({\it i.e.}, $r$-process) elements in
old Galactic halo and globular cluster stars.
Recent  observations of the $r$-process-enriched star
\bd\
include new abundance determinations for
the neutron-capture elements Cd I (Z=48), Lu II (Z = 71) and
Os II (Z = 76),
the first detections of these elements in metal-poor $r$-process-enriched
halo stars.
Combining these and previous observations, we have now detected
32 $n$-capture elements in \bd.  This is the most of any
metal-poor halo star to date.
For the most $r$-process-rich  ({\it i.e.} [Eu/Fe] $\simeq$ 1) halo stars, 
such as CS 22892-052
and \bd,
abundance comparisons show that the  
heaviest
stable $n$-capture elements ({\it i.e.}, Ba and above, Z $\geq$ 56) are
consistent with a scaled solar system $r$-process abundance
distribution. The lighter $n$-capture element abundances in
these stars, however, do
not conform to the solar pattern.
These comparisons, as well as recent observations of
heavy elements in metal-poor globular clusters, suggest
the possibility of multiple synthesis mechanisms
for the $n$-capture elements.
The heavy element abundance patterns in most metal-poor halo stars
do not resemble that of CS 22892-052,
but the presence of heavy elements
such as Ba 
in nearly all metal-poor stars
without $s$-process enrichment
indicates that $r$-process enrichment 
in the early Galaxy is common.

\end{abstract}

\clearpage 

\section{Introduction}

Members of our group have been involved in long-term studies of abundances in  
Galactic halo stars. 
These studies have been designed to address a number of important 
issues, including: the synthesis mechanisms of 
the heavy,
specifically, neutron capture ($n$-capture) 
elements, early in the history of the
Galaxy; the identities of the earliest stellar generations,
the progenitors of the
halo stars; the site or sites for the synthesis of the 
rapid $n$-capture ({\it i.e.}, $r$-process) material
throughout the Galaxy; the Galactic Chemical Evolution (GCE) of
the elements; and by employing the abundances of the
radioactive elements (Th and U) as 
chronometers, the ages of the oldest stars, and hence the lower limit
on the age of the Galaxy and the Universe. (See \citealt{truran02},
\citealt{snedencowan03}, \citealt{cowan04}, and \citealt{sneden08}
for discussions of these related and significant topics.)  

In the following paper we review some of the results of our studies,
starting with new stellar abundance determinations arising from
more accurate laboratory atomic data 
in \S II, followed by  abundance comparisons
of the
lighter and heavier $n$-capture elements in the $r$-process-rich stars 
in \S III, with new species  detections in the star \bd \
and the ubiquitous nature of the $r$-process throughout the
Galaxy described in  
sections \S IV 
and \S V, respectively. We end with our Conclusions in \S VI.

\section{Atomic Data Improvements and Abundance Determinations}

Stellar abundance determinations of the $n$-capture elements in 
Galactic halo stars have become increasingly more accurate 
over the last decade with typical errors now 
of less than 10\% \citep{sneden08}.
Much of that improvement in the precision of the stellar abundances 
has been due to increasingly more accurate laboratory atomic data. 
New measurements of the transition probabilities  
have been published  for the rare earth elements (REE) and several
others, including:
La~II \citep{lawler01a};
Ce~II (\citealt{palmeri00};
   and recently transition probabilities for 921 lines for Ce~II,
   \citealt{lawler09});
Pr~II \citep{ivarsson01};
Nd~II (transition probabilities for  more than 700
   Nd~II lines, \citealt{denhartog03});
Sm~II (\citealt{xu03};
   and recently transition probabilities for more
   than 900 Sm~II lines, \citealt{lawler06});
Eu~I, II, and III (\citealt{lawler01c}; \citealt{denhartog02});
Gd~II \citep{denhartog06};
Tb~II (\citealt{denhartog01}; \citealt{lawler01b});
Dy~I and II \citep{wickliffe00};
Ho~II \citep{lawler04};
Er~II (transition probabilities for 418
lines of Er II, \citealt{lawler08});
Tm~I and II (\citealt{anderson96}; \citealt{wickliffe97});
Lu~I, II, and III (\citealt{denhartog98}; \citealt{quinet99}, 
   \citealt{fedchak00});
Hf~II \citep{lawler07};
Os~I and II (\citealt{ivarsson03,ivarsson04}; \citealt{quinet06});
Ir~I and II (\citealt{ivarsson03,ivarsson04}; \citealt{xu07});
Pt~I \citep{denhartog05};
Au~I and II (\citealt{fivet06}; \citealt{biemont07});
Pb~I \citep{biemont00};
Th~II \citep{nilsson02a};
U~II \citep{nilsson02b};
and finally  in new, more precise solar and stellar abundances of
Pr, Dy, Tm, Yb, and Lu \citep{sneden09}.
 
\begin{figure*}
\plotone{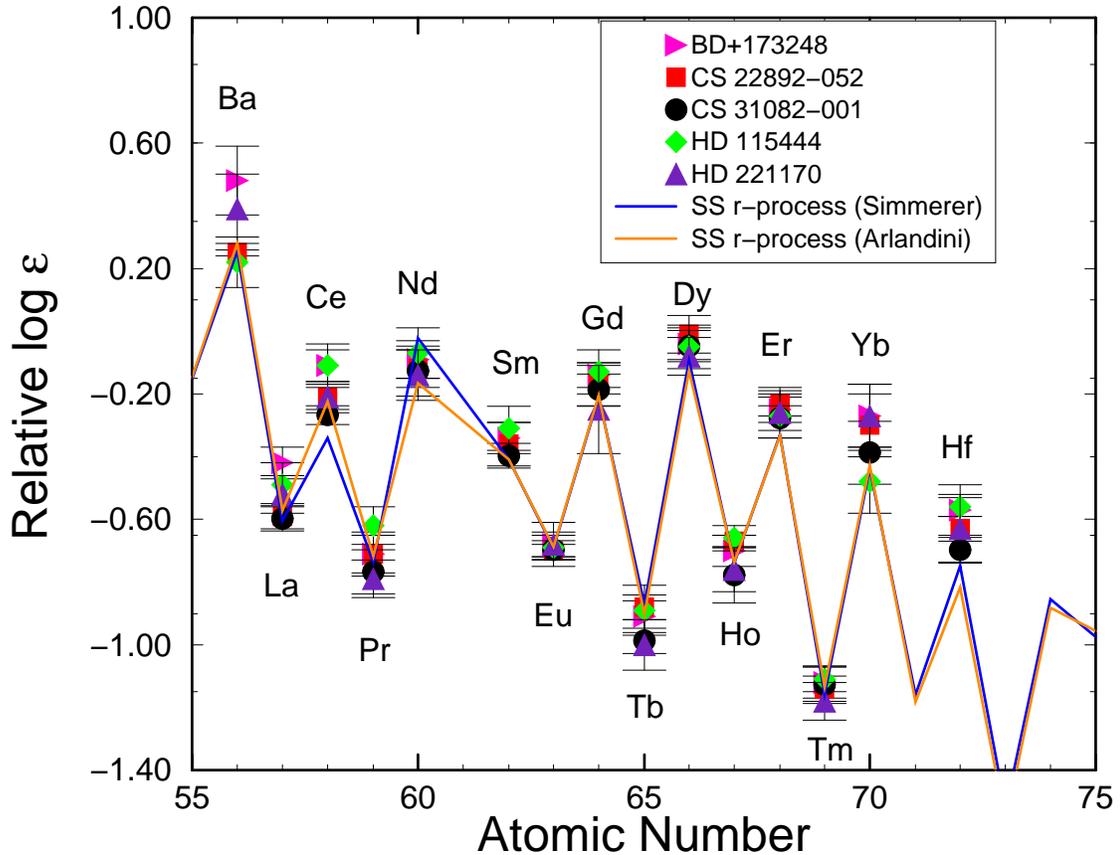}
\caption{Recent
abundance determinations in five $r$-process rich 
stars, based upon
new atomic lab data, compared with two solar system $r$-process only 
predictions.  The abundances in each star have been normalized to the
element Eu. After \citet{sneden09}.
Reproduced by permission of the AAS.
}
\label{f8}
\end{figure*}

These new atomic data have been employed to redetermine the solar 
and stellar abundances.
We show in  Figure~\ref{f8} (from \citealt{sneden09})
the relative REE, and Hf,   
abundances in five $r$-process rich stars:  \bd, \cs, \mbox{CS~31082-001},
HD~115444 and HD~221170, 
where   
the abundance distributions have been scaled to the element Eu for
these comparisons. Also shown in Figure~\ref{f8}
are two  Solar System $r$-process-only 
abundance predictions from \citet{arlandini99} (based upon a 
stellar model calculation)  and 
\citet{simmerer04}   (based upon the ``classical'' $r$-process residual 
method) that are also matched to the Eu abundances.  
What is clear from the figure is that all of the
REE abundances---as well as Hf, which is a heavier interpeak element---are 
in the same relative proportions from 
star-to-star and with respect to the solar $r$-process abundances.
This agreement between the heavier $n$-capture elements and the
Solar System $r$-process abundance distribution 
has been noted in the past (see, {\it e.g.}, \citealt{sneden03}), but   
the overall agreement has become much more
precise, and convincing,  as a result of the new atomic laboratory data.

\section{Abundance Comparisons}

We can also compare more comprehensive---not just the 
REE---elemental abundance determinations for the
$r$-process-rich halo stars. This is potentially
a more rewarding enterprise, as it can illuminate the complex
nucleosynthetic
origin of the lightest $n$-capture elements, and can provide new ways of
looking at the age of the Galactic halo. 

\subsection{Heavy $n$-capture Elements}

\begin{figure*}
    \centering
\includegraphics[angle=-90,width=7.00in]{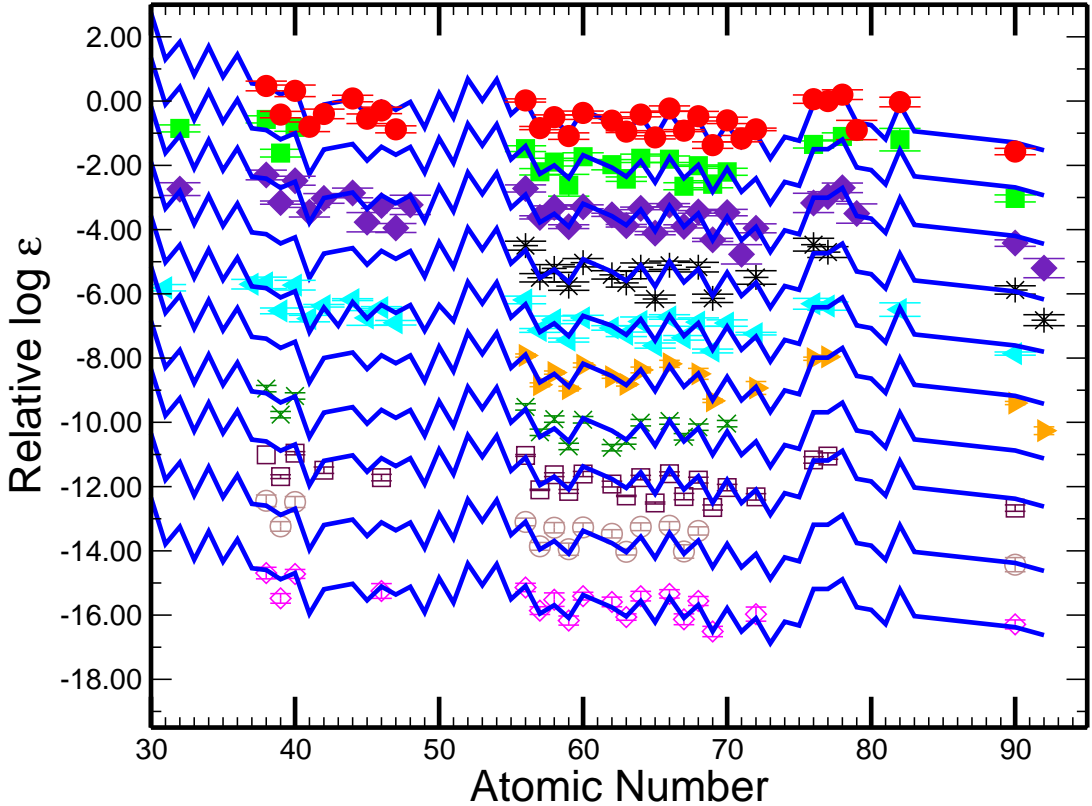}
\caption{
Abundance comparisons between 10 $r$-process rich stars and the Solar 
System $r$-process values.
See text for references.  
Adapted from  \citet{sneden11}.
}
\label{compnew4}
\end{figure*}

We show in Figure~\ref{compnew4} abundance comparisons with extensive 
elemental  data for
10 $r$-process-rich stars
(from the top): 
filled (red) circles, CS~22892-052 \citep{sneden03,sneden09};
filled (green) squares, HD~115444 \citep{westin00,sneden09,hansen11};
filled (purple) diamonds, \bd\ \citep{cowan02,roederer10b};
(black) stars, CS~31082-001 \citep{hill02,plez04};
solid (turquoise) left-pointing triangles, HD~221170 \citep{ivans06,sneden09};
solid (orange) right-pointing triangles, HE~1523-0901 \citep{frebel07};
(green) crosses, CS~22953-003 \citep{francois07}; 
open (maroon) squares, HE~2327-5642 \citep{mashonkina10};
open (brown) circles, CS~29491-069 \citep{hayek09}; and
open (magenta) triangles, HE~1219-0312 \citep{hayek09}.
The abundances of all the stars except 
\cs\ have been vertically displaced
downwards for display purposes.
In each case the 
solid lines are (scaled) solar system $r$-process only predictions from
\citet{simmerer04} that have been matched to the Eu abundances.

The figure indicates that for the  ten stars plotted, the
abundances of {\it all} of the heavier stable $n$-capture elements 
({\it i.e.}, Ba and above) are 
consistent with the relative solar system $r$-process abundance distribution 
(see also \citealt{sneden09}).
Earlier work had demonstrated this agreement for 
several $r$-process rich stars (where [Eu/Fe] $\simeq$ 1), including \cs, 
and the addition of still more such $r$-process-rich stars supports that
conclusion.

\subsection{Light $n$-capture Elements}

While the heavier $n$-capture elements appear to be consistent with 
the scaled solar system $r$-process curve, the lighter $n$-capture 
elements (Z $<$ 56) seem to fall below that same solar curve. 
One problem in analyzing this region of interest 
is that there have been relatively few stellar observations of these
lighter $n$-capture elements until now.
With the limited amount of data it is not yet clear if the pattern
is the same from star-to-star for the lighter $n$-capture elements in 
these $r$-process rich stars. 

There has been extensive work on trying to understand the 
synthesis of these elements.
Observations of 4 metal-poor $r$-enriched stars
by \citet{crawford98}
suggested that Ag (Z = 47) was produced in rough proportion
to the heavy elements in stars with $-$2.2~$<$~[Fe/H]~$< -$~1.2.
\citet{wasserburg96} and \citet{mcwilliam98} pointed out
that multiple sources of heavy elements (other than the $s$-process)
were required to account for the observed abundances
in the solar system and extremely metal-poor stars, respectively.
\citet{travaglio04} quantized this effect, noting that Sr-Zr 
Solar System abundances
could not be totally accounted for from traditional sources, such as 
the $r$-process, the (main) $s$-process and the weak $s$-process. 
They suggested that
the remaining (missing) abundances---8\% for Sr to 18\%  
for Y and Zr---came from
a light element primary process (LEPP).   
Travaglio et al.\ also noted,
``The discrepancy in the $r$-fraction of Sr-Y-Zr between the
$r$-residuals method and the \cs\ abundances
becomes even larger for elements from Ru to Cd: the weak
$s$-process does not contribute to elements from Ru to Cd. As
noted [previously], this discrepancy suggests an even
more complex multisource nucleosynthetic origin for elements
like Ru, Rh, Pd, Ag, and Cd.''

\citet{montes07} extended studies of the LEPP and suggested
that a range of $n$-capture elements, perhaps even including heavier 
elements such as Ba,  might have a contribution
from this primary process. (Since, however, Ba in $r$-process 
rich stars is consistent with  
the solar $r$-process abundances, such contributions  
for these heavier 
elements must be quite small.)
They noted, in particular, that this LEPP might
have been important in  synthesizing the abundances in the $r$-process poor
star HD~122563.    

Further insight into the (complicated) 
origin of the lighter $n$-capture elements is 
provided by the detections of Ge (Z = 32) in a few stars.
\citet{cowan05} noted a correlation of Ge with the
iron abundances in the halo
stars with $-$3.0~$\lesssim$~[Fe/H]~$\lesssim -$1.5,
suggesting that the Ge is being produced along with the Fe-group elements 
at these low metallicities.
To produce the protons needed to satisfy 
such a correlation, a new neutrino ({\it i.e.}, $\nu$-p) process that might 
occur in supernovae was suggested \citep{frohlich06}.
We note that for higher ({\it i.e.}, at  solar) 
metallicities,  Ge is considered a neutron-capture element, 
synthesized in the $r$-process (52\%) and the $s$-process (48\%) 
(Simmerer et al. 2004; Sneden et al. 2008). Thus, there should be 
a change in the slope  of the Ge abundances from low metallicities to
higher metallicities, a behavior  that has not yet been observed.

\begin{figure*}
\centering
\plotone{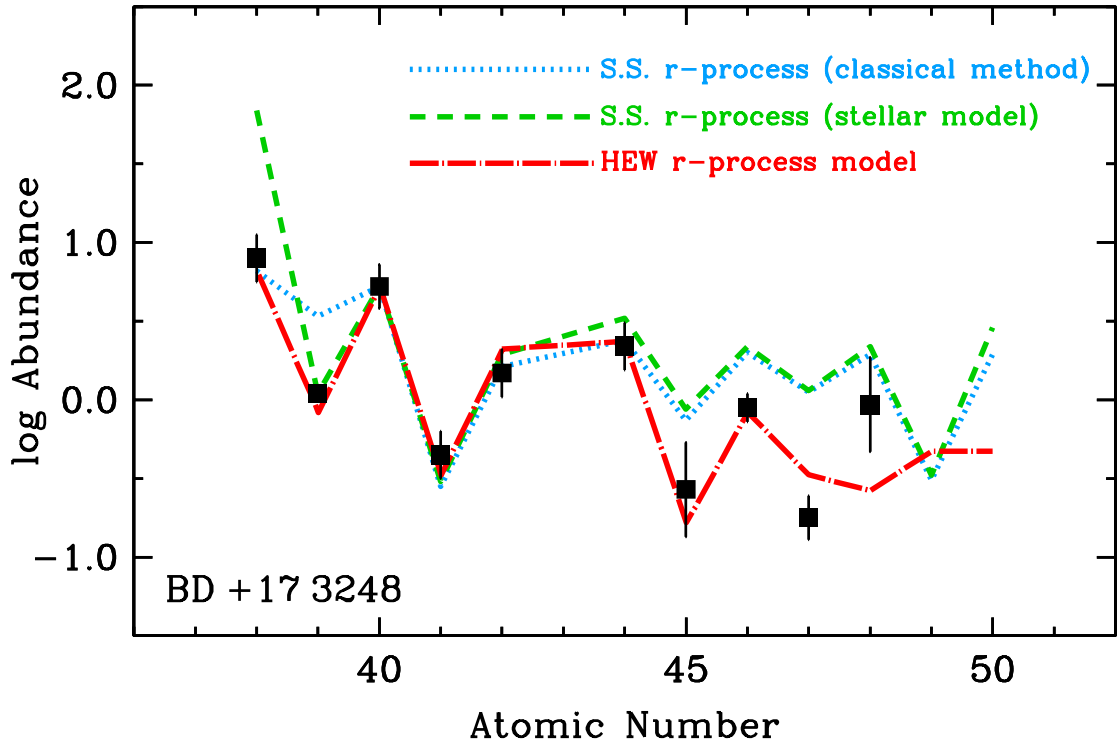}
\caption{$r$-process abundance predictions for light $n$-capture elements
compared with observations of  \bd \ from \citet{roederer10b}.
See text for further details.} 
\label{bdp17}
\end{figure*}

We show in  
Figure~\ref{bdp17} several $r$-process predictions for the 
lighter $n$-capture element abundances compared with observations of 
those elements in \bd \ from \citet{roederer10b}.
The two Solar System $r$-process models (``classical'' and ``stellar model'') 
reproduce some of these elements but begin to diverge from the 
observed abundances at Rh (Z = 45). 
Also shown in 
Figure~\ref{bdp17} are predictions from 
a High Entropy Wind (HEW) model, that might be
typical in a core-collapse (or Type II) supernova  (\citealt{farouqi09};
K.-L.~Kratz, private communication.)
This model gives a better fit to the abundances, but does not reproduce 
the observed odd-even effects in Ag (Z = 47)and Cd (Z = 48) in this star
(resembling a trend discovered in other 
$r$-enriched stars by \citealt{johnson02}).
Recent work by \citet{hansen11} to study Pd (Z = 46) and Ag 
abundances in stars with $-$3.2~$\lesssim$~[Fe/H]~$\lesssim -$0.6
confirms the divergence between observations and simulation predictions.

These comparisons between calculations and observations do in fact 
argue for a combination of processes to reproduce 
the observed stellar abundances of some of these light $n$-capture elements. 
This combination of processes might include (contributions from) 
the main $r$-process, the LEPP, the $\nu$-p process,
charged-particle reactions  
accompanied by $\beta$-delayed fission 
and the weak $r$-process ({\it e.g.}, \citealt{kratz07}).
(See, {\it e.g.}, 
\citealt{farouqi09,farouqi10}, \citealt{roederer10a,roederer10b},
and \citealt{arcones11}
for further discussion.)
It may also be that during the synthesis 
the main $r$-process  and the LEPP are separate processes,
and that the abundance patterns in all metal-poor stars could be
reproduced by mixing their yields \citep{montes07}.
Alternatively, it may be  
that the $r$-process and the LEPP
can be produced in the same events, but sometimes
only the lower neutron density components  are present 
\citep{kratz07,farouqi09}.
It has also been suggested that the  
heavier and lighter 
$n$-capture elements are synthesized in separate sites (see {\it e.g.}, 
\citealt{qian08}). 

New observations of heavy elements in metal-poor globular cluster
stars reaffirm the abundance patterns seen in field stars.
In the globular cluster M92, \citet{roederer11b} found that the 
observed star-to-star dispersion in Y (Z = 39) and Zr (Z = 40)
is the same as for the Fe-group elements ({\it i.e.}, consistent
with observational uncertainty only).
Yet, the Ba \citep{sneden00}, La, Eu, and Ho abundances exhibit 
significantly larger star-to-star dispersion that cannot be
attributed to observational uncertainty alone.
Furthermore, the Ba and heavier elements were produced by $r$-process
nucleosynthesis without any $s$-process contributions.
This indicates that, as in the field stars, 
these two groups of elements could not have
formed entirely in the same nucleosynthetic process in M92.

\section{New Species Detections}

\begin{figure}
\centering
\plotone{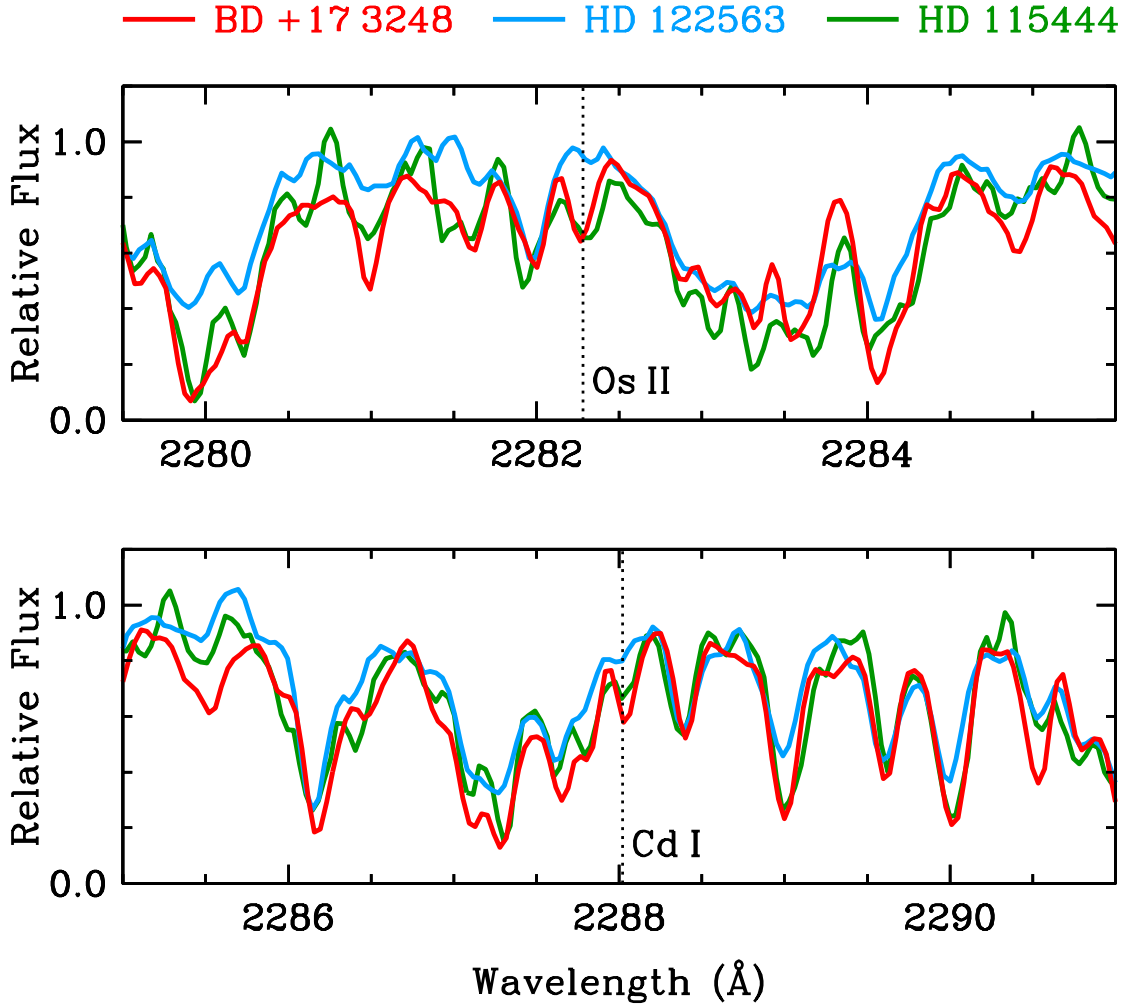}
\vskip0pt
\caption{
HST (near-UV) 
spectral regions containing Os~II and Cd~I lines  in \bd,  HD~122563,
and HD~115444 from \citet{roederer10b}.
Reproduced by permission of the AAS.
}
\label{f1}
\end{figure}

\citet{roederer10b} reanalyzed near-UV spectra obtained with
HST/STIS of the 
star \bd.  
(See also \citealt{cowan02,cowan05} for earlier HST observations of \bd.)
We show in 
Figure~\ref{f1} (from \citealt{roederer10b})
spectral regions around Os~II and Cd~I lines in the
stars \bd, HD~122563 and HD~115444.  There is a clear detection of Os~II
in both \bd\ and HD~115444 but not in HD~122563. The star HD~115444 is 
similar in metallicity and atmospheric parameters to HD~122563 
(see \citealt{westin00}), but much more 
$r$-process rich: [Eu/Fe] = 0.7 versus $-$0.5, respectively. 
In the lower panel of 
Figure~\ref{f1} we see the presence of Cd~I in \bd \ and HD~115444, as well
as a weak detection in HD~122563. 
Synthetic fits to these spectra in \bd\ and HD~122563 
indicate the presence of 
Cd~I and Lu~II lines in both stars, 
as well as the detection (and upper limit of) Os~II in the same 
two stars, respectively.  
This work was 
the first to detect Cd~I, Lu~II, and Os~II
in metal-poor halo stars.

\begin{figure*}
    \centering
\vskip 0.55in
\includegraphics[width=5.00in]{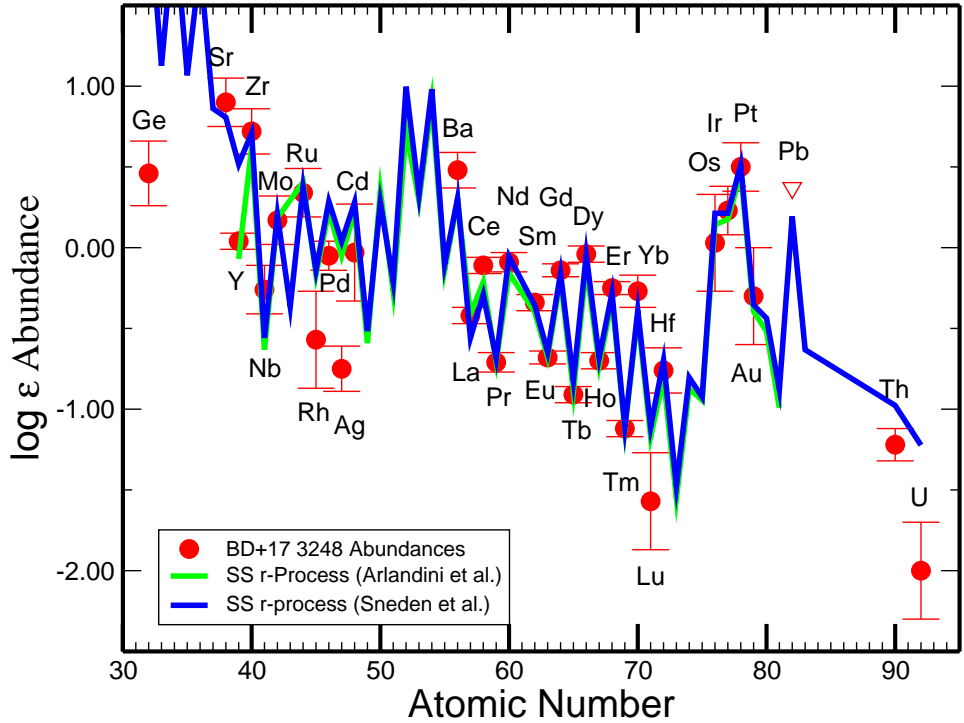}
\vskip 0.35in
\caption{
The total observed abundance distribution in \bd. 
There are a total of 32---not including Ge---detections of $n$-capture elements,
the most in any metal-poor halo star. This distribution is 
compared with the
two Solar System $r$-process curves from \citet{simmerer04} and
\citet{arlandini99}.
}
\label{bdfourthb}
\end{figure*}

In addition to these new detections,  
\citet{roederer10b} employed Keck/HIRES spectra   
to derive new abundances of Mo I, Ru I and Rh I in this star.  
Combining these abundance determinations led to the detection of a total of 
32 $n$-capture species  ---the most of any metal-poor halo star.
(Previously, CS~22892-052 had the most such detections.)
Further, we note that 
the total detections in \bd \ did not count the element Ge. 
And while Ge may be
synthesized in proton-rich processes early in the history of the
Galaxy,  
it is classified as a $n$-capture element in Solar System material
(see \citealt{simmerer04} and \citealt{cowan05}).
We illustrate this total abundance distribution 
in Figure~\ref{bdfourthb} compared with the 
two Solar System $r$-process curves from \citet{simmerer04} and
\citet{arlandini99}. We again see the close agreement between the heavier
$n$-capture elements and (both of) the predictions for  
the Solar System $r$-process curve, as well as   
the deviation between the abundances of the 
lighter $n$-capture elements and that 
same curve.

\section{The $r$-process Throughout the Galaxy}

\begin{figure*}
    \centering
\includegraphics[width=5.45in]{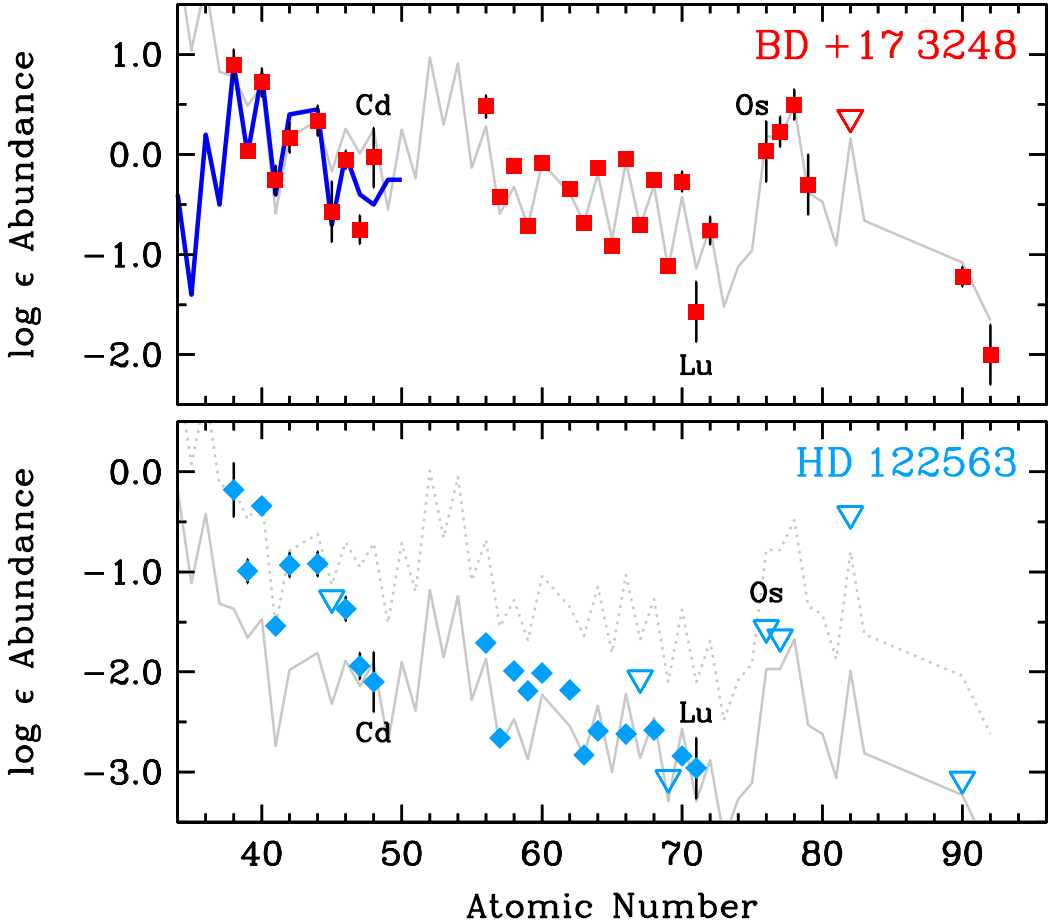}
\caption{
Abundance distributions in \bd\ and HD~122563 with detections indicated by
filled symbols and upper limits by downward-pointing open triangles.
The new measurements of Os, Cd, and Lu illustrated in  Figure~\ref{f1} 
are labeled.
In the top panel (\bd) 
the bold curve is an HEW calculation from \citet{farouqi09}
normalized to Sr, while the solid line is the Solar System $r$-process 
curve \citep{sneden08} normalized to Eu. 
In the bottom panel (HD~122563) the solar curve is normalized both to Eu 
(solid line) and
Sr (dotted line). 
Abundances were obtained from \citet{cowan02,cowan05}, \citet{honda06},
\citet{roederer09,roederer10b}, and \citet{sneden09}.
Figure from \citet{roederer10b}.
Reproduced by permission of the AAS.
}
\label{f4}
\end{figure*}

The results of \citet{roederer10b} also confirm earlier work indicating
significant  differences in the abundances between $r$-process rich stars, such
as \bd, and $r$-process poor stars, such as HD~122563. 
This difference is shown clearly in Figure~\ref{f4}. The abundance 
distribution for \bd \ (shown in the top panel) is relatively flat---compare 
the abundance of Sr with Ba---and
is consistent with the scaled Solar System $r$-process abundances for 
the heavy $n$-capture elements.  
In contrast the lower panel of this figure indicates that the abundances
in the $r$-process poor HD~122563 fall off dramatically with increasing
atomic number---again compare the abundance of Sr with Ba.    

  \begin{figure*}
    \centering
\includegraphics[angle=-90,width=6.00in]{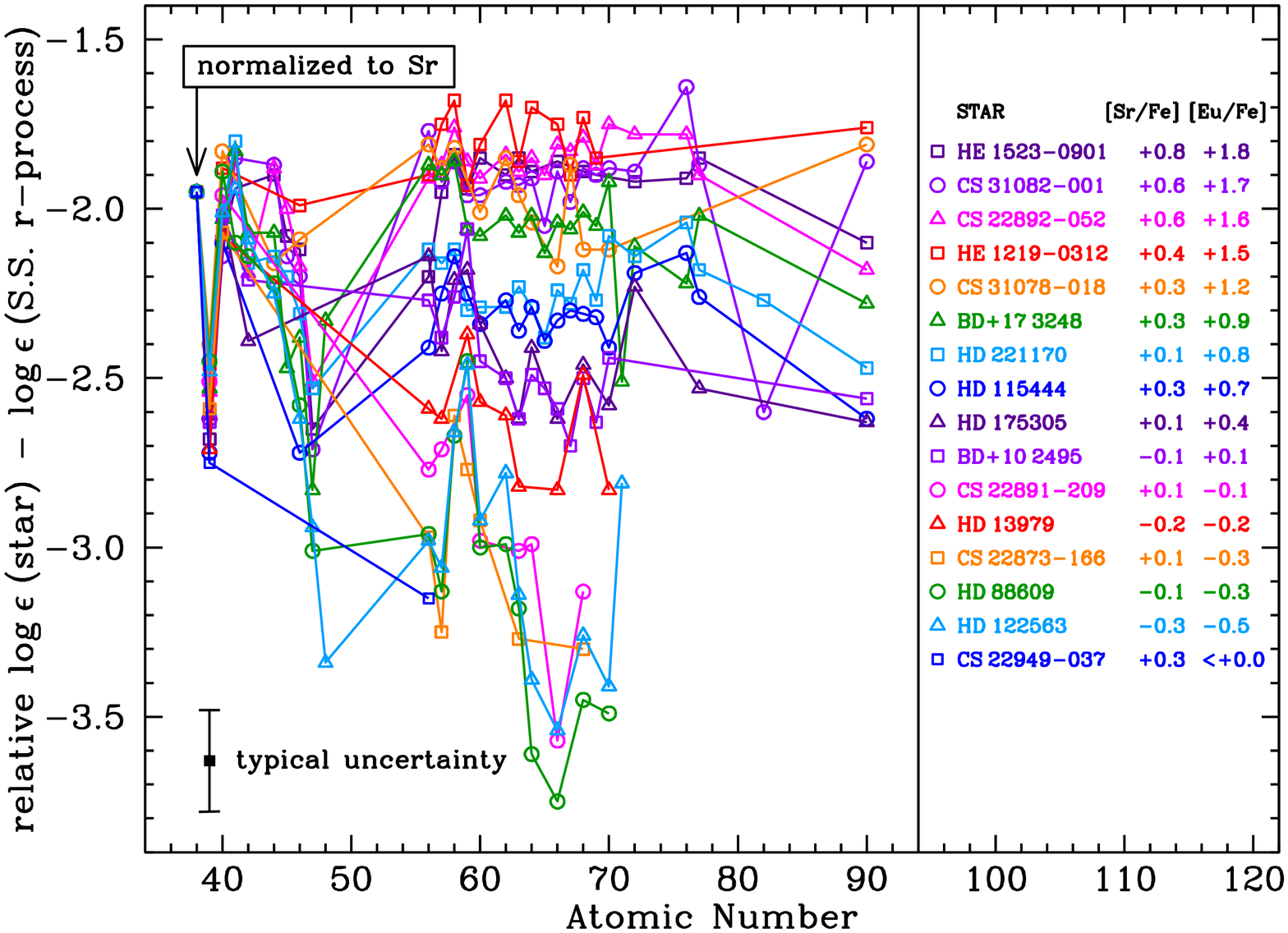}
\caption{ 
Differences between Solar System $r$-process abundances and stellar abundances
for 16 metal-poor stars, normalized to Sr. 
The stars are listed in order of descending [Eu/Fe], and that value and
[Sr/Fe] are listed in the box to the  right in the figure. 
A value for a typical uncertainty is illustrated in the lower left.
Note the difference in the abundance pattern between the $r$-process 
rich star \cs\ and that of HD~122563, with the other stars falling 
between those extremes.
Abundance references are as follows:
S.S. $r$-process  abundances \citep{sneden08};
\mbox{HE~1523-0901} (\citealt{frebel07} and A.\ Frebel, 
2009, private communication);
\mbox{CS~31082-001} \citep{hill02,plez04,sneden09};
\mbox{CS~22892-052} \citep{sneden03,sneden09};
\mbox{HE~1219-0312} \citep{hayek09,roederer09};
\mbox{UMi-COS~82} \citep{aoki07};
\mbox{CS~31078-018} \citep{lai08};
\mbox{CS~30306-132} \citep{honda04};
\mbox{BD$+$17~3248} \citep{cowan02,sneden09,roederer10b};
\mbox{HD~221170} \citep{ivans06,sneden09};
\mbox{HD~115444} \citep{westin00,roederer09,sneden09};
\mbox{HD~175305} \citep{roederer10c};
\mbox{BD$+$29~2356} \citep{roederer10c};
\mbox{BD$+$10~2495} \citep{roederer10c};
\mbox{CS~22891-209} \citep{francois07};
\mbox{HD~128279} \citep{roederer10c};
\mbox{HD~13979} (I.\ Roederer et al., in preparation);
\mbox{CS~29518-051} \citep{francois07};
\mbox{CS~22873-166} \citep{francois07};
\mbox{HD~88609} \citep{honda07};
\mbox{CS~29491-053} \citep{francois07};
\mbox{HD~122563} \citep{honda06,roederer10b}; and
\mbox{CS~22949-037} \citep{depagne02}.
Figure from \citet{roederer10a}.
Reproduced by permission of the AAS.
}
\label{f11b}
\end{figure*}

It is clear from much work 
({\it e.g.}, Honda et al. 2006, 2007) that the abundances even in a 
star such as HD~122563 do come from the 
$r$-process---the source for the $s$-process, 
low- or intermediate-mass stars on the AGB with longer evolutionary
timescales, have not had sufficient time to evolve prior to the formation
of this metal-poor halo star (cf.\ \citealt{truran81}).
Instead,  one can think of the abundance distribution 
in HD~122563,  illustrated in Figure~\ref{f4},  as the result of 
an ``incomplete $r$-process''---there were not sufficient numbers of 
neutrons to form all of the heavier $n$-capture elements, 
particularly the ``third-peak'' elements of Os, Ir, and Pt. 
In the classical ``waiting point approximation'' the
lighter $n$-capture elements are synthesized from lower neutron number 
density (n$_n$)
fluxes, typically 10$^{20}$--10$^{24}$,  with the heavier $n$-capture elements 
(and the total $r$-process abundance distribution) requiring 
values of n$_n$ = 10$^{23}$--10$^{28}$ cm$^{-3}$ 
(see Figures 5 and 6 of \citealt{kratz07}).
Physically in this ``incomplete'' or ``weak $r$-process,''
the neutron flux was too low to push the $r$-process ``path''
far enough away from the valley of $\beta$-stability 
to reach the higher mass numbers after 
$\alpha$ and $\beta$-decays back to stable nuclides.
Instead the lower neutron number densities result in the 
$r$-process path being too close to the valley of stability 
leading to a diminution in the
abundances of the heavier $n$-capture elements. 
The lighter $n$-capture elements, such as Sr, in this star 
may have formed as a result of this incomplete or 
weak $r$-process, or the LEPP,  or combinations as described previously for the 
$r$-process rich stars.

This analysis was extended to a larger sample by \citet{roederer10a}
and is illustrated in Figure~\ref{f11b}.
We show the differences between the abundance distributions of 16  
metal-poor stars, normalized to Sr, 
compared with the Solar System $r$-process distribution \citep{sneden08}. 
The stars are plotted in order of descending values of
[Eu/Fe], a measure of their $r$-process richness. Thus, we see near the 
top \cs\ with a value of [Eu/Fe] = 1.6 and near the bottom, HD~122563 with
[Eu/Fe] = $-$0.5. The figure illustrates 
 the relative flatness of the distributions of 
the most $r$-process-rich stars ([Eu/Fe] $\simeq$ 1) 
with respect to the solar curves, while the
$r$-process poor stars have abundances that fall off sharply with 
increasing atomic number. 
It is also clear from Figure~\ref{f11b} 
that there are a range of abundance distributions
falling between 
these two extreme examples.  
(We note that Figure~\ref{f11b} should not be taken as an unbiased 
distribution of stars at low metallicity.)

We emphasize four important points here. 
First, not all of the metal-poor stars  have the same abundance pattern
as \cs, only those that are $r$-process rich. 
Second, while the distributions
are different between the $r$-process rich and poor stars there 
is no indication of $s$-process synthesis for these elements. 
Thus, all of the
elements in these stars were synthesized in the $r$-process, at least for
the heavier $n$-capture elements, and $r$-process material was common in the
early Galaxy.  
Third, the approximate downward displacement from the top to the bottom
of the Figure~\ref{f11b} (a measure of the decreasing [Eu/Sr] ratio) 
roughly scales as the [Eu/Fe] ratio, listed
in the right-hand panel.
This can be understood as follows: since the abundance patterns are
normalized to Sr, and if Sr is roughly proportional to Fe in these stars
(with a moderate degree of scatter---cf.\ Figure~7 of \citealt{roederer10a}),
then {\it of course} the [Eu/Sr] ratio roughly follows [Eu/Fe].
(See also \citealt{aoki05}.)
Finally, we note that Ba has been detected in all of these stars 
and the vast majority of low-metallicity field and globular cluster stars
studied to date.
Only in a few Local Group dwarf galaxies do Ba upper limits
hint that Ba (and, by inference, all heavier elements) may be
extremely deficient or absent
\citep{fulbright04,koch08,frebel10}.


\section{Conclusions}

Extensive studies have demonstrated the presence of 
$n$-capture elements in the atmospheres of
metal-poor halo and globular cluster stars. 
New detections of the $n$-capture elements Cd~I (Z = 48), Lu~II 
(Z = 71) and Os~II (Z = 76),
derived from HST/STIS spectra, 
have been made in several metal-poor halo stars. 
These were the first detections of these species in such stars.
Supplementing these observations with Keck data 
and new measurements of  Mo I, Ru I and Rh I,  
we reported the detections of 32 $n$-capture 
elements in \bd. This is currently the most detections of these 
elements in any metal-poor halo star, supplanting the previous ``champion''
\cs. 

Comparisons  among the most 
$r$-process-rich stars ([Eu/Fe] $\simeq$ 1) demonstrate that  the heaver  
stable elements (from Ba and above) are remarkably consistent from star-to-star 
 and consistent with the (scaled) solar system $r$-process distribution.
Detailed comparisons of the REE (along with Hf)  among 
a well-studied group of $r$-process-rich stars,  employing new
experimental atomic data, 
strongly supports this finding.
The newly determined, and lab-based, stellar abundances  are more
precise and show very little scatter from star-to-star and with
respect to the Solar System $r$-process abundances.
This suggests that the $r$-process produced these elements 
early in the history of the Galaxy and that the same type(s) of process
was responsible for the synthesis of the $r$-process elements at the 
time of the formation of the Solar System.

While the heavier elements appear to have formed from the main 
$r$-process and are
apparently consistent with the solar $r$-process abundances, 
the lighter $n$-capture element abundances in
these stars do
not conform to the solar pattern.
There have been little data in these stars 
until recently, but now with the new detections 
of Cd and increasing Pd and Ag detections, some patterns are becoming clear.
First, the main $r$-process alone is not responsible for the synthesis of
these lighter $n$-capture elements. Instead, other processes, alone or in
combination, may have responsible for such formation.
These processes include a so-called 
``weak'' $r$-process (with lower values of n$_n$), the LEPP, 
the $\nu$-p process, or charged particle 
reactions in the HEW of a core-collapse supernova.  
It is also not clear whether different processes are responsible for 
different mass regions with one for Ge, a different one for Sr-Zr and 
still another for Pd, Ag, and Cd.
It is also not clear whether these processes operate separately from 
each other or in the same site, or whether different mass ranges of 
the $n$-capture elements are synthesized in different sites.
Clearly, much more work needs to be undertaken to understand the
formation of these lighter $n$-capture elements.

The stellar abundance signatures of the heaviest of these  elements, 
i.e., Ba and above, are consistent with the rapid neutron
capture process, $r$-process, but not the $s$-process in these 
old stars. Similar conclusions are found for stars in the ancient globular 
clusters 
with comparable abundance spreads in the $r$-process elements
(see, {\it e.g.}, \citealt{gratton04}, \citealt{sobeck11}, 
\citealt{roederer11a}, and \citealt{roederer11b}).
There is also a clear distinction between the abundance patterns of the 
$r$-process rich stars such as \cs\ and the $r$-process poor stars like
HD~122563. The latter seem to have an element pattern that was 
formed as a result of a ``weak'' or ``incomplete'' $r$-process.
Most of the old, metal-poor halo stars have abundance distributions
that fall between the extremes of \cs\ and HD~122563. 
However, the very presence of  
$n$-capture elements in the spectra of these stars  
argues for $r$-process events being a common occurrence 
early in the 
history of
the Galaxy. 

Finally, we note the need for additional stellar observations, 
particularly of the UV regions of the spectra {\it only} accessible
using STIS or COS aboard HST.   
These observations require 
high signal-to-noise ratios and high resolution to identify faint lines 
in crowded spectral regions. Also we will require more laboratory 
atomic data for 
elements that have not been well studied to improve the precision of 
the stellar and solar abundances. 
Additional experimental nuclear data,
not yet available, 
 for the heaviest neutron-rich nuclei 
that participate in the $r$-process,  
will be critical to these studies. Until that time 
new, more physically based,
 theoretical prescriptions for nuclear masses, half-lives, etc.\
for these $r$-process nuclei will be necessary.
New theoretical models of supernova explosions and detailed synthesis
scenarios, such as might occur in the HEW, will be very important to help to
identify the site or sites 
for the $r$-process, a search that has been ongoing since 1957.

\section{Acknowledgments}
We thank our colleagues for all of their contributions and helpful
discussions. We particularly are grateful for all of the 
contributions from George W.\ Preston,  as he celebrates his
80th birthday. Partial scientific support for this research was 
provided by the NSF (grants AST~07-07447 to J.J.C., AST~09-08978 to C.S.,
and AST~09-07732 to J.E.L.).  
I.U.R.\ is supported by the Carnegie Institution of Washington
through the Carnegie Observatories Fellowship.

\end{document}